\documentclass[letter]{stile/aa}

\usepackage{txfonts}

\usepackage[english]{babel}
\usepackage{amsmath,amssymb}
\usepackage{graphicx, subfig}
\usepackage[normalem]{ulem}

\usepackage{verbatim}

\usepackage{multirow}
\usepackage{mathtools}
\usepackage{epstopdf}

\usepackage{url}
\usepackage{xcolor,color}
\usepackage{orcidlink}

\usepackage{natbib,twoopt}
\usepackage[hyphenbreaks]{breakurl}

\bibpunct{(}{)}{;}{a}{}{,}             
\definecolor{cobalt}{rgb}{0.06, 0.2, 0.65}
\hypersetup{
  colorlinks,
  citecolor=cobalt,
  linkcolor=[rgb]{0.8, 0.2, 1.0},
  urlcolor=cobalt,
}
\makeatletter
  \newcommandtwoopt{\citeads}[3][][]{\href{http://adsabs.harvard.edu/abs/#3}%
    {\def\hyper@linkstart##1##2{}%
     \let\hyper@linkend\@empty\citealp[#1][#2]{#3}}}
  \newcommandtwoopt{\citepads}[3][][]{\href{http://adsabs.harvard.edu/abs/#3}%
    {\def\hyper@linkstart##1##2{}%
     \let\hyper@linkend\@empty\citep[#1][#2]{#3}}}
  \newcommandtwoopt{\citetads}[3][][]{\href{http://adsabs.harvard.edu/abs/#3}%
    {\def\hyper@linkstart##1##2{}%
     \let\hyper@linkend\@empty\citet[#1][#2]{#3}}}
  \newcommandtwoopt{\citeyearads}[3][][]%
    {\href{http://adsabs.harvard.edu/abs/#3}
    {\def\hyper@linkstart##1##2{}%
     \let\hyper@linkend\@empty\citeyear[#1][#2]{#3}}}
\makeatother

\newcommand\code[1]{\textsc{\MakeLowercase{#1}}}
\newcommand{\quotes}[1]{``#1''}

\def\msun{{\rm M}_{\odot}}
\def\zsun{{\rm Z}_{\odot}}

\def\dsun{\mathcal{D}_{\odot}}

\def\angstrom{\textrm{A\kern -1.3ex\raisebox{0.6ex}{$^\circ$}}}

\def\myr{{\rm Myr}}

\def\msunyr{\msun\,{\rm yr}^{-1}}

\def\dust{\mathcal{D}}

\def\be{\begin{equation}} 
\def\ee{\end{equation}} 
\def\ba{\begin{eqnarray}} 
\def\ea{\end{eqnarray}} 
\def\gtsima{$\; \buildrel > \over \sim \;$}
\def\ltsima{$\; \buildrel < \over \sim \;$}
\def\gsim{\lower.5ex\hbox{\gtsima}} 
\def\lsim{\lower.5ex\hbox{\ltsima}}
\def\prosima{$\; \buildrel \propto \over \sim \;$} 
\def\simgt{\lower.5ex\hbox{\gtsima}} 
\def\simlt{\lower.5ex\hbox{\ltsima}} 
\def\simpr{\lower.5ex\hbox{\prosima}}

\definecolor{mkcolor}{HTML}{01abdf}
\definecolor{apcolor}{HTML}{b3003b}
\definecolor{afcolor}{HTML}{01bdff}
\definecolor{lvcolor}{HTML}{ff9933}
\definecolor{colorecommenta}{HTML}{9a24d1}

\graphicspath{{plots/}}

\begin{document}

\title{Stochastic star formation in early galaxies: JWST implications}
\titlerunning{SF variability}
\author{
  A. Pallottini \orcidlink{0000-0002-7129-5761} \inst{1}\fnmsep\thanks{\href{mailto:andrea.pallottini@sns.it}{andrea.pallottini@sns.it}}
  \and A. Ferrara \orcidlink{0000-0002-9400-7312} \inst{1}
     }
\authorrunning{Pallottini \& Ferrara}
\institute{Scuola Normale Superiore, Piazza dei Cavalieri 7, 56126 Pisa, Italy}
\date{Received July 6, 2023; accepted 14 August, 2023}

 
\abstract
{The star formation rate (SFR) in high redshift galaxies is expected to be time-variable due to competing physical processes. Such stochastic variability might boost the luminosity of galaxies, possibly explaining the over-abundance seen at $z\gsim 10$ by JWST.
}
{We aim at quantifying the amplitude and timescales of such variability, and identifying the key driving physical processes.}
{We select 245 $z=7.7$ galaxies with stellar mass $5\times 10^{6}\lsim M_\star/\msun\lsim 5\times 10^{10}$ from \code{SERRA}, a suite of high-resolution, radiation-hydrodynamic cosmological simulations. After fitting the average SFR trend, $\langle {\rm SFR} \rangle$, we quantify the time-dependent variation, $\delta(t) \equiv \log [\rm SFR/\langle {\rm SFR} \rangle]$ for each system, and perform a periodogram analysis to search for periodicity modulations.}
{We find that $\delta(t)$ is distributed as a zero-mean Gaussian, with standard deviation $\sigma_\delta \simeq 0.24$ (corresponding to a UV magnitude s.d. $\sigma_{\rm UV} \simeq 0.61$) that is independent of $M_\star$. However, the modulation timescale increases with stellar mass: $t_\delta \sim (9, 50, 100)\, \rm Myr$ for $M_\star \sim  (0.1, 1, 5)\times 10^9\, \msun$, respectively. These timescales are imprinted on the SFR by different processes: (i) photoevaporation, (ii) supernova explosions, and (iii) cosmological accretion/merging dominating in low, intermediate, and high mass systems, respectively.
}
{The predicted SFR variations cannot account for the required $z\gsim 10$ UV luminosity function boost. Other processes, such as radiation-driven outflows clearing the dust, must then be invoked to explain the enhanced luminosity of super-early systems.}
\maketitle
%

\section{Introduction}
Understanding the physics regulating the evolution of star formation in galaxies is among the most fundamental problems in present-day cosmology and astrophysics.

Despite the variety of non-linear processes involved, the huge dynamical range ($10^3 - 10^{11}\msun$), and the diversities of environment probed (from single molecular clouds to very massive galaxies), the star formation rate (SFR) seems to depend on a deceptively simple and universal function of the gas mass \citep{schmidt:1959,kennicutt:1998,krumholz:2012_b}.
On top of that, in the relatively low redshift Universe ($z\lsim 4$), remarkably accurate predictions for the galaxy mass build-up can be obtained from minimal, quasi-equilibrium models, where the star formation history (SFH) is a consequence of a \quotes{bath-tube} balance between cosmic accretion and stellar/QSO feedback \citep{bouche:2010,dekel:2014}. These models manage to explain fundamental observables such as, e.g., the mass-metallicity relation \citep{maiolino:2018}, or the dust content \citep{dayal:2014}, and are able to predict the ultraviolet luminosity functions (UV LF) up to $z\sim 6$ \citep{tacchella:2018}.

However, as we consider progressively higher-redshift galaxies, the timescale of the feedback processes (e.g. energy input from massive stars and supernovae) becomes comparable or longer than the dynamical time of the system \citep{faucherguiger:2018}, thus decreasing and delaying their regulatory effect on the SFR. As a consequence, SFR variations develop a stochastic character \citep[][]{orr:2019}.
Stated differently, the quasi-equilibrium assumption should eventually break down at a sufficiently high $z$: early galaxies are then expected to be bursty, a feature highly influencing the determination of their properties \citep{furlanetto:2022}. Indeed, already at $z\sim 5-8$, the burstiness/suppression of SFR of galaxies, i.e. being above/below the \citet{schmidt:1959}-\citet{kennicutt:1998} relation, is key to explain \citep{ferrara:2019,pallottini:2019} the deviation from [CII]-SFR relation observed in local ($z=0$) galaxies \citep{de_looze:2014,herrera-camus:2018}.
The natural expectation is that such a stochastic variability should have an even more prominent impact for the super-early galaxies seen by JWST.

Indeed, a stochastic SFR might affect the observable fraction of high-$z$ galaxies, as they flicker in and out of the current magnitude-limited surveys \citep{sun:2023}. Moreover, it seems that more abrupt variations with respect to the relatively gentle ones induced by supernova explosions are needed \citep{gelli:2023,dome:2023} to explain the detection of the rapid quenching observed in some early systems \citep{looser:2023}.

Perhaps more importantly, stochastic variability is among the possible mechanisms invoked to explain the over abundance of $z\gsim 10$ galaxies that has been probed via the UV luminosity function inferred from the public JWST Early Release Science programs \citep{borsani:2022,finkelstein:2022_a,naidu:2022,adams:2023,atek:2023,donnan:2023,harikane:2023}, and the results from the CEERS \citep{finkelstein:2022} and GLASS \citep{castellano:2022,treu:2022,santini:2023} surveys.
Pre-JWST models cannot predict the observed flatness and limited evolution of the bright-end of the LF. Thus, multiple cosmological and astrophysical scenarios have been proposed to explain such a conundrum.

On the one hand, there are rather extreme frameworks, e.g. (a) requiring $\Lambda$CDM modifications invoking a different primordial power spectrum \citep{liu:2022,padmanabhan:2023}, or (b) suggesting a feedback-free boosted star formation \citep{dekel:2023,qin:2023}.
On the other hand, more mundane options boost the UV luminosity function via (c) the temporary removal of dust \citep{ferrara:2023} as a consequence of radiation-driven outflows \citep{ziparo:2023} promoted by high specific SFR and small galactic sizes \citep{fiore:2023} or (d) stochastic variability of the SFR \citep{mason:2023,shen:2023,mirocha:2023,munoz:2023}.

The latter scenario can be formulated as follows. Time-variance in the dark matter (DM) halo assembly history induces a stochastic variation in gas mass that can accrete on a galaxy, thus causing a flickering of the SFR \citep{mason:2023}. Such flickering shifts low mass galaxies to brighter-than-expected UV luminosities, in principle compensating for the overabundance seen at $z\gsim 10$ by JWST if the distribution of the UV variability has a standard deviation (s.d.) of $\sigma_{\rm UV} \simgt 2$ \citep{shen:2023}.
However, this interpretation tends to be in contrast with the high stellar masses derived from Spectral Energy Density (SED) fitting of JWST observations \citep{santini:2023}. However, as the SED determination of $M_\star$ is degenerate with dust attenuation \citep{markov:2023}, the \quotes{bursty} scenario could still be viable provided low mass galaxies are substantially reddened \citep{mirocha:2023}.

In principle, the stochastic UV boost is agnostic with respect to the physical mechanism driving it \citep{shen:2023}. In fact, UV variability can be induced by the DM halo assembly history \citep{mason:2023}, an unbalance in the feedback regulation of SFR \citep{mirocha:2023}, and, to some extent, it includes the effect of variation caused by a reduced dust attenuation \citep{ferrara:2023}.
In this \textit{Letter}, we aim at clarifying the amplitude of the SFR variability by using the \code{serra} simulations \citep[][]{pallottini:2022}.

\section{Method}

\begin{figure*}
\centering
\includegraphics[width=0.49\textwidth]{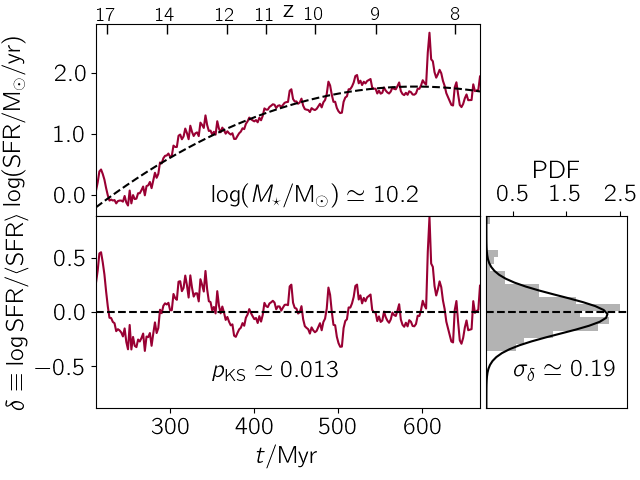}
\includegraphics[width=0.49\textwidth]{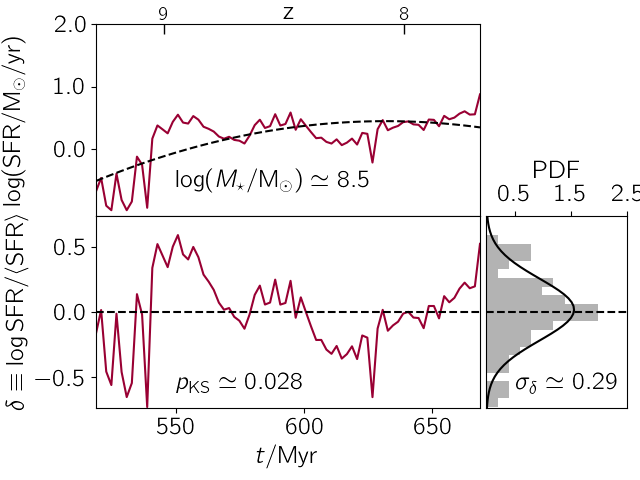}
\caption{Example of SFH fitting and variability definitions for high ($\log M_\star /\msun \simeq 10.2$, {\bf left panel}) and low mass ($\log M_\star /\msun \simeq 8.5$, {\bf right panel}) galaxies in the \code{SERRA} simulation at $z=7.7$.
For each galaxy, in the upper panel we plot the star formation ($\rm SFR$) as a function of cosmic time ($t$) and its fit ($\langle \rm SFR \rangle$, eq. \ref{eq:definition_sfr_fit}) as a continuous and dashed line, respectively.
Note that we report the SFR starting from the time when the galaxy has a stellar mass higher then $10^{6.5}\msun$, the SFR is averaged on temporal bins of $2\,\rm Myr$, and in the upper axis we plot redshift ($z$).
In the lower left panel, we plot the variation ($\delta$, eq. \ref{eq:eq:definition_variation}) as a function of time, add a constant line for no variation ($\delta =0$) to guide the eye, and report the $p$-value from the Kolmogorov-Smirnov ($p_{\rm KS}$) for the fit.
In the lower right inset, we show the Probability Distribution Function (${\rm PDF}$) of the variation, along with its Gaussian fit and standard deviation ($\sigma_\delta$).
\label{fig:fit_examples}}
\end{figure*}

\code{SERRA} is a suite of cosmological zoom-in simulations that follows the evolution of $z\gsim 6$ galaxies \citep[][]{pallottini:2022}.
DM, gas, stars are evolved with a customized version of adaptive mesh refinement code \code{RAMSES} \citep{teyssier:2002}.
\code{KROME} \citep{grassi:2014} is used to generate a procedural numerical solver \citep[cfr.][]{branca:2023} to follow the non-equilibrium chemistry of H, H$^{+}$, H$^{-}$, He, He$^{+}$, He$^{++}$, H$_{2}$, H$_{2}^{+}$ and electrons \citep{bovino:2016,pallottini:2017_b}.

Metallicity ($Z$) is tracked as the sum of heavy elements, assuming solar abundance ratios of different metal species \citep{asplund:2009}; we assume that dust follows metals by adopting a constant dust-to-metal ratio $\dust = \dsun (Z/\zsun)$ -- where $\dsun/\zsun \simeq 0.3$ for the MW \citep[][]{hirashita:2002} and a MW-like grain size distribution \citep{weingartner:2001}.
An initial $Z=10^{-3}\zsun$ metallicity floor is adopted, as expected from a pre-enrichment scenario \citep[][]{wise:2012, pallottini:2014}.

Radiation is tracked on-the-fly using the moment-based radiative transfer code \code{RAMSES-RT} \citep{rosdahl:2013}, that is coupled to the chemical evolution of the gas \citep{pallottini:2019,decataldo:2019}; the energy bins cover part of the Habing band (1 bin, $6.0<{h}\nu <11.2$), the Lyman-Werner band (1 bin, $11.2<{h}\nu <13.6$) to account for H$_2$ photoevaporation, and the ionisation of H up to the first ionisation level of He (3 bins, $13.6<{h}\nu <24.59$).

Using a \citet{schmidt:1959}-\citet{kennicutt:1998} like relation, H$_2$ is converted into stars, which act as a source for mechanical energy, photons, and reprocessed elements, depending on the metallicity $Z_\star$ and the age $t_\star$ of the stellar population \citep{bertelli:1994}.
Feedback includes supernovae, winds from massive stars, and an approximate treatment of radiation pressure\footnote{Note we do not adopt the standard radiation pressure prescription of \code{ramses-rt} \citep[][]{rosdahl:2015} as the feedback modelling was ported from simulations that were not including radiative transfer \citep[see][for details]{pallottini:2019}.}: depending on the kind, the energy input can be both thermal and kinetic, and we account for the dissipation of energy in molecular clouds for supernova (SN) blastwaves \citet{pallottini:2017}.

Simulations are initialised at $z=100$ from cosmological initial conditions generated with \code{music} \citep{hahn:2011} by adopting a \emph{Planck} $\Lambda$CDM model\footnote{$\Lambda$CDM model with vacuum, matter, and baryon densities in units of the critical density $\Omega_{\Lambda}= 0.692$, $\Omega_{m}= 0.308$, $\Omega_{b}= 0.0481$, Hubble constant $\rm H_0=67.8\, {\rm km}\,{\rm s}^{-1}\,{\rm Mpc}^{-1}$, spectral index $n=0.967$, and $\sigma_{8}=0.826$ \citep[][]{planck_collaboration:2014}.}.
Each simulation zoom-in on a target DM halo ($M_h\sim 10^{12}\msun$ at $z = 6$) and its environment ($\sim (2\,{\rm Mpc}/{\rm h})^{3}$, i.e. about 23 times the Lagrangian volume of the target halo), reaching a gas mass and spatial resolution of $\Delta m_g \simeq 1.2\times 10^4 \msun$ and $\Delta l \simeq 21\,{\rm pc}$ at $z=7.7$ in the densest regions, i.e. typical mass and size of Galactic molecular clouds.

\section{Analysis}

%
We use the $z=7.7$ sample of \code{serra} galaxies with $M_\star>10^{6.5}\msun$, i.e. those containing $\gsim 100$ star particles, for a total of $245$ objects. The maximum stellar mass of the sample is about $5\times 10^{10} \msun$.
In the original \citet[][]{pallottini:2022}, 202 galaxies were presented, here we add 43 new simulated objects, that share the same physical modelling but are obtained for different seeds for the perturbations of the cosmological initial conditions\footnote{Within two virial radii all galaxies in the sample have a contamination of $\lsim 0.1\%$ of low resolution particles outside the zoom-in region.}.

\subsection{Star formation: stochastic variability}

%
High-$z$ galaxies are observed to have an increasing SFH \citep[e.g.][]{topping:2022} that is predicted to be exponential in our simulations \citep[i.e.][in particular see Fig. 2]{pallottini:2017_b}.
Thus, for each galaxy, we reconstruct the SFH with a $2\,\myr$ time resolution, and define its average trend by adopting a polynomial fit in log space:
\begin{equation}\label{eq:definition_sfr_fit}
\log \langle {\rm SFR}/\msunyr \rangle \equiv \sum_{i=0}^2 p_i \left(\frac{t}{\rm Myr}\right)^{i}\,,
\end{equation}
where the order is limited to the $2^{\rm nd}$ to avoid removing oscillatory terms that are possibly present in the SFR (see Sec. \ref{sec:analysis_periodicity} for the \textit{a posteriori} motivation); alternative methods that can be used to pinpoint the variability also implied informed choices, e.g. the maximum order selected for a principal component analysis \citep{montero:2021} and the number and width of time windows for SFR averaging for non-parametric fits \citep{leja:2019}. We consider the SFH of a galaxy starting from the time $t_0$ when the stellar mass is larger than $10^{6.5} \msun$.
Using $\langle {\rm SFR} \rangle$ from eq. \ref{eq:definition_sfr_fit}, we define the stochastic time variability (or flickering) of the star formation as the residual of the fit:
\begin{equation}\label{eq:eq:definition_variation} 
\delta \equiv \log \frac{\rm SFR}{\langle {\rm SFR} \rangle}\,.
\end{equation}

%
Fig. \ref{fig:fit_examples} shows an example of the procedure for two representative galaxies with high ($\log M_\star /\msun \simeq 10.2$) and low ($\log M_\star /\msun \simeq 8.5$) mass.
As expected for high-redshift systems \citep[see, e.g. Fig. 3 in][]{pallottini:2022}, \code{serra} galaxies show a time-increasing SFR. The most massive galaxy has a higher SFR, with peaks up to $200\,\msunyr$, and a longer SFH, typically starting at $z\sim 20$ and lasting for about $500\, \myr$. The lower mass galaxy barely reaches $10\,\msunyr$, and it forms stars only for a short time span ($\sim 150\, \myr$).
In both cases, the average trend is well captured by the fitting procedure, as can be appreciated by eye and as highlighted by the low $p$-value of the two Kolmogorov-Smirnov (KS) tests performed on $ {\rm SFR}$ and $\langle {\rm SFR} \rangle$. 

For both galaxies, $\delta$ is roughly distributed as a zero-mean Gaussian; this is partially expected, considering the goodness of the fit ($p$-value $\ll 1$), and the fact the flickering is defined as a residual (eq. \ref{eq:eq:definition_variation}). However, the amplitude of the s.d. is similar for the two galaxies, $\sigma_\delta\simeq 0.29$ (0.19) for the low (high) mass galaxy. In both cases the maximum amplitude of the flickering is $|\delta|_{\rm max}\sim 0.7-0.8$.
However, such extreme fluctuations are rare and short-lived, e.g. the high mass galaxy exhibits a $\simeq 4.7\,\sigma_\delta$ peak (probability $\simeq 2\times 10^{-4}\%$) at $t\simeq 600\,\myr$ which lasts for $\sim 10\,\myr$ only.
Further, due to the Gaussian nature of the fluctuations, the duty cycle associated to episodes of mini-quenching \citep{dome:2023} of amplitude $\nu \sigma_\delta$ can be written as $f_{\rm duty}(\nu \sigma_\delta) = {\rm erf}(\nu)$, with $\rm erf$ being the error function.

%
The same fitting procedure can be applied to the full sample of 245 galaxies. After checking that for all objects the fit gives a satisfactorily low $p$-value\footnote{Adopting a $1^{\rm st}$ order for the fit (eq. \ref{eq:definition_sfr_fit}) yields a worst KS performance but qualitatively the same results presented in Sec. \ref{sec:analysis_periodicity}.} ($\leq 0.05$), we collect $\delta(t) - M_\star(t)$ pairs in each $2\,\myr$ time bin. Treating each pair independently from the SFH of the original galaxy, we can investigate how the flickering depend on the stellar mass.

\begin{figure}
\centering
\includegraphics[width=0.49\textwidth]{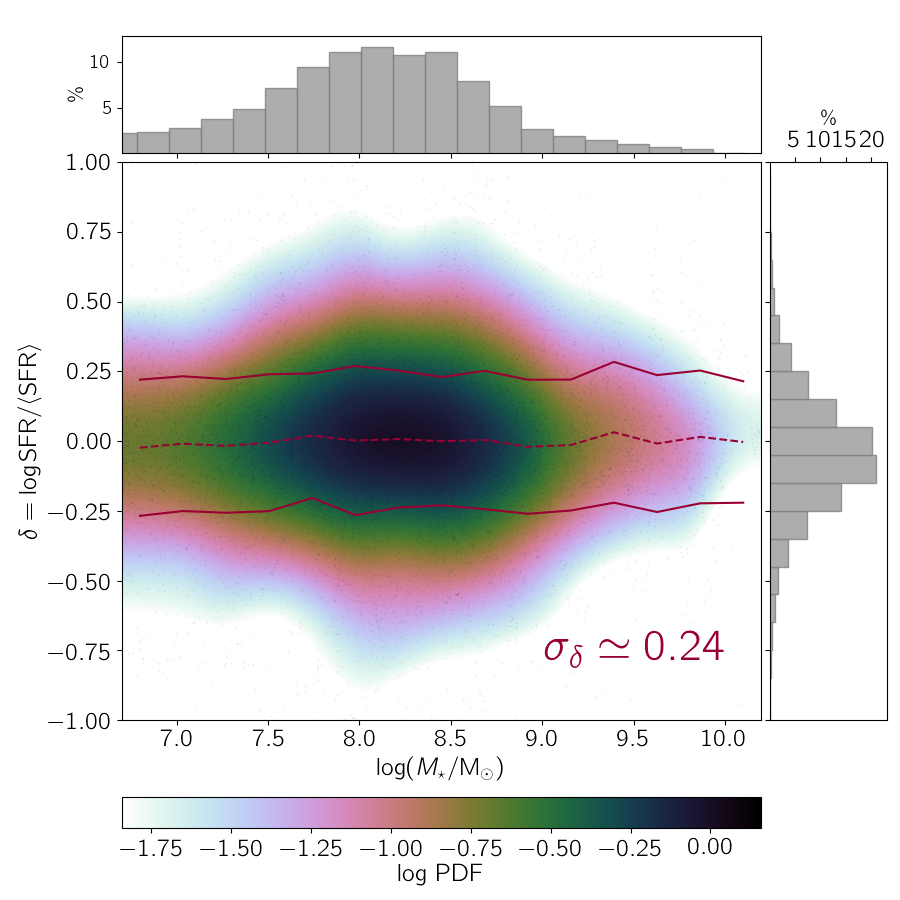}
\caption{
Distribution of the SFR variation ($\delta$) as a function of stellar mass ($M_\star$).
The colourbar indicates the PDF in the $\delta$-$M_\star$ plane, with dashed and solid lines we report the mean and variance of $\delta$ as a function of $M_\star$.
The average standard deviation is indicated as an inset.
Horizontal and vertical insets are the 1-D PDFs, normalised so to give the probability in each bin.
\label{fig:variability}
}
\end{figure}

The result of the analysis is reported in Fig. \ref{fig:variability}, where we show the Probability Distribution Function (PDF) in the $\delta$-$M_\star$ plane.
The most striking feature is that the standard deviation of the flickering shows a flat trend, i.e. $\sigma_\delta\simeq 0.24$ almost independently of $M_\star$. Interestingly, the scatter of the $z<6$ galaxy Main Sequence \citep{popesso:2023} is also constant across the mass range $10^{8.5-11.5}\, \msun$, albeit with a smaller s.d. of $0.09$. 
Higher deviations ($|\delta|_{\rm max}\simeq 0.75 \simeq 3.1 \sigma_\delta$) can occur in $7.5 \lsim \log (M_\star/\msun) \lsim 8.5$ objects. However, such extreme values are not statistically relevant (probability $\simeq 0.2\%$), and they do not affect the value of $\sigma_\delta$. 

Similarly to \citet{furlanetto:2022}, we find that flickering is common in all galaxies, up to the massive ones hosted by DM halos $M_{h}\simeq 1.2\times 10^{12}\msun$.  However, while these authors find that $\delta(M_\star)$ slightly decreases with mass, as in their case modulation is solely induced by the delay in the feedback regulation of the SFR, in \code{serra} the SFR variability can be additionally caused by cosmic accretion and/or merging events \citep[for an analysis, see][]{kohandel:2020} which play an important role in massive galaxies. We will return on this point in Sec. \ref{sec:analysis_periodicity}.

Finally, the $\delta$-PDF is symmetric, possibly suggesting that intense phases of SFR activity (high $\delta$) are followed by quiescent phases (low $\delta$): as a galaxy enters in a starburst regime, the enhanced mechanical, radiative, and turbulent feedback energy injection can temporarily quench or reduce the SFR \citep{looser:2023,gelli:2023}. Gas cooling is likely to play a major role in re-enabling a stable SFR after burst; this study is left to future work \citet{gelli:2023b}.

\subsection{SFR variability: periodicity analysis}\label{sec:analysis_periodicity}


\begin{figure}
\centering
\includegraphics[width=0.49\textwidth]{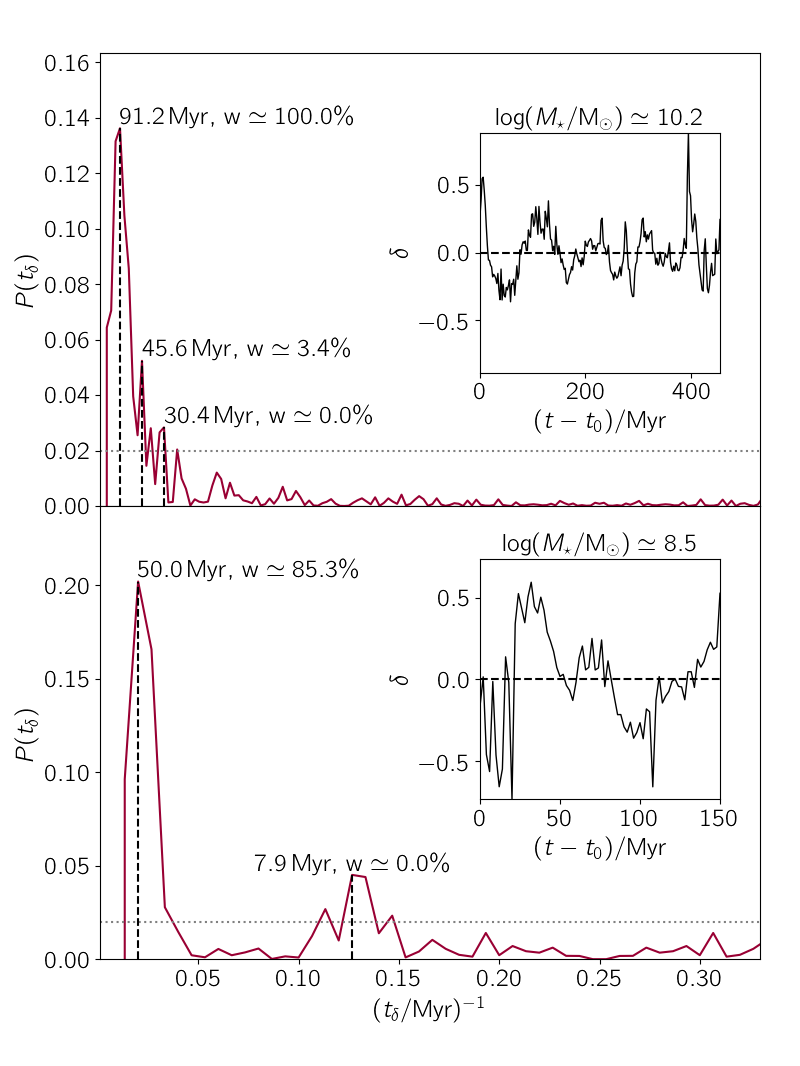}
\caption{
Example of periodicity analysis of the SFR variability for the same high ({\bf upper panel}) and low ({\bf lower}) mass galaxies presented in Fig. \ref{fig:fit_examples}.
For each galaxy, we plot the \citet{lomb:1976}-\citet{scargle:1998} periodogram of the variability ($P(t_\delta)$) as a function of the frequency ($t_\delta^{-1}$). 
%
Each peak recovered above the selected noise threshold (dotted line) is marked in the plot with a dashed line and its characteristic time ($t_\delta$) and significance ($w$) are given.
As a reference, in the inset we plot the evolution of $\delta$ vs cosmic time ($t$) shifted by $t_0$, i.e. the time when the stellar mass becomes $>10^{6.5} \msun$.
\label{fig:periodogram_examples}}
\end{figure}

\begin{figure}
\centering
\includegraphics[width=0.49\textwidth]{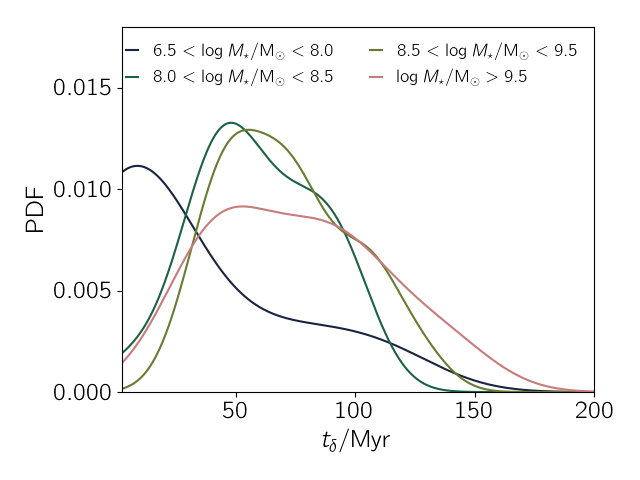}
\caption{Distributions of characteristic timescales ($t_\delta$) of the periodicity of the flickering $\delta$.
Each line is a PDF computed weighting for the peak significance (w) for a sub-sample of galaxies falling in a different mass by $z=7.7$; for increasing mass bins, the sub-samples contain 35, 89, 100, and 11 galaxies (typically $\sim 4$ peaks per galaxy are found). The PDF is computed via a kernel density estimator, adopting \citet{silverman:1986}'s rule of thumb for the bandwidth size. 
\label{fig:variability:timing}
}
\end{figure}

%
To investigate the presence of periodicity in $\delta(t)$, we adopt the \citet{lomb:1976}-\citet{scargle:1998} periodogram (see \citealt{vanderplas:2018} for a practical guide). Instead of a Fourier power spectrum analysis, we use a periodogram estimate, since it yields a cleaner peak recognition, and a more convenient way to define the peak significance via the false alarm probability i.e. $w = 1-{\rm false\, alarm}$.

For illustration, in Fig. \ref{fig:periodogram_examples} we show the periodogram analysis\footnote{We extract all the peaks above a selected noise threshold. The results are mostly unchanged when the noise threshold is varied within reasonable limits.} for the same two  galaxies used in Fig. \ref{fig:fit_examples}.
For the $\log M_\star /\msun \simeq 10.2$ galaxy, the periodogram identifies a prominent (significance $w\simeq 100\%$) peak with characteristic time scale $t_\delta\simeq 91.2\,\myr$. Such periodicity is visible by eye in the SFH of the galaxy, and corresponds to a modulation consistent with cosmological accretion/merging timescales of massive ($M_h \sim 10^{11} \msun$) DM halos at $z\simeq6-10$ \citep[][Fig. 1]{furlanetto:2017}.
Further, a less significant ($w\simeq 3.4\%$) peak is found at $t_\delta\simeq 45.6\,\myr$, corresponding to the time at which all massive stars ($8\lsim m_\star/\msun \lsim 40$) have exploded as SN. Finally, a $t_\delta\simeq 30.4\,\myr$ feature is also present. Although possibly associated with SN feedback as well, the peak has a very low significance ($w\simeq 0.0\%$), and we reject it as spurious.

For the $\log M_\star /\msun \simeq 8.5$ galaxy, the periodogram shows a solid ($w\simeq 85.3\%$) peak with a SN-compatible timescale ($t_\delta\simeq 50.0\,\myr$).
A second peak is present in the periodogram of this galaxy at $t_\delta\simeq 7.9\,\myr$. This timescale is consistent with radiative feedback from massive stars. UV radiation might efficiently and quickly affect a low mass, dust-poor galaxy by photo-dissociating $\rm H_2$, thus temporarily halting the star formation \citep[for an extreme example, see the Alyssum effect in][]{pallottini:2022}. In practice, though, the significance of the peak is almost null, hence we cannot firmly confirm our hypothesis in this case. 
Note that higher order fits (eq.  \ref{eq:definition_sfr_fit}) tend to suppress longer timescale modulations present in the SFH, e.g. the significant $t_\delta\simeq 50.0\,\myr$ ($t_\delta\simeq 91.2\,\myr$) for the $\log M_\star /\msun \simeq 8.5$ ($\log M_\star /\msun \simeq 10.2$) galaxy.

%
To complete our study, we perform the periodogram analysis on all our galaxies.
Based on the value of their stellar mass at $z=7.7$, we divide the galaxies in 4 sub-samples with increasing $M_\star$, i.e. $6.5 < \log(M_\star/\msun) < 8$, $8 < \log(M_\star/\msun) < 8.5$, $8.5 < \log(M_\star/\msun) < 9.5$, $\log(M_\star/\msun) > 9.5$, which contain 35, 89, 100, and 11 galaxies, respectively.
For each galaxy, we collect all the $t_\delta$ peaks found above the noise threshold (usually $\sim 4$ peaks per galaxy are found), and compute the PDF of the timescales of the sub-sample by weighing each $t_\delta$ with its significance $w$. The results are shown in Fig. \ref{fig:variability:timing}.

The low mass sub-sample ($6.5 < \log(M_\star/\msun) < 8$) shows a clear maximum at $t_\delta\sim 9 \,\myr$ and has a long tail, extending up to $\gsim 100 \,\myr$.
For intermediate mass systems, the maximum of the PDF shifts at $48$ and $55\,\myr$ for the $8 < \log(M_\star/\msun) < 8.5$ and $8.5 < \log(M_\star/\msun) < 9.5$ sub-samples, respectively; both distributions retain their high-$t_\delta$ tails, which become more pronounced. 
For the most massive galaxies ($\log(M_\star/\msun) > 9.5$), the maximum remains around the same timescales, but the tail has grown so much in importance that the PDF is almost flat between the maximum at $t_\delta \sim 53\,\myr$ and up to $t_\delta\sim 100\,\myr$.

Physically, this can be interpreted as follows. Galaxies in early stages of growth ($M_\star/\msun \leq 10^8 \msun$) have their SFR modulated by radiative feedback effects (i.e. $\rm H_2$ photo-dissociation). As they grow more massive ($M_\star \sim 10^9\msun$), the timescale for the SFR stochasticity is dominated by SN; for galaxies in the high-mass end ($M_\star/\msun\gsim 5\times 10^{9}$), SN co-regulate the flickering along with cosmological accretion/merging, that is equally important but acts on longer timescales.

\subsection{Flickering: implications for the UV luminosity}


\citet{shen:2023} note that $\sigma_{\rm UV}$ can be decomposed in contributions from variations produced by accretion history, delay in the feedback regulation of the star formation, and dust attenuation.
However, in our analysis we have seen that the first two contributions are both encapsulated in $\delta$, i.e. the flickering in \code{serra} galaxies is mostly induced by feedback regulation, with DM assembly history playing a role at the high-mass end.

To discuss how the flickering impacts the bright end of the luminosity function, it is sufficient to assume that the UV luminosity is sensitive to the instantaneous SFR
\begin{equation}\label{eq:photon_production}
M_{\rm UV} = -2.5 \log(\rm SFR \times \rm constant) + M_0\,,
\end{equation}
where the constant accounts for the efficiency of photo production depending on the stellar population \citep[e.g.][for a review]{madau:2014} and $M_0$ is a normalisation constant.
As noted in \citet[][]{furlanetto:2022}, the $\rm UV$ luminosity is sensitive to the SFR in the last $\sim 20\,\myr$, thus if the timescale for the stochastic variation is shorter, the induced UV variation should be reduced. As shown in Sec. \ref{sec:analysis_periodicity}, $M_\star\gsim 10^8\msun$ galaxies indeed have $t_\delta\gsim 50\,\myr$ (Fig. \ref{fig:variability:timing}); note that assuming an instantaneous photon production (eq. \ref{eq:photon_production}) maximises the UV variation. While in principle the photo production depends on $Z_\star$, \code{serra} galaxies quickly reach close to solar values \citep{gelli:2020}, thus for $M_{\rm UV}$ the induced variation is negligible.

Thus, considering the instantaneous UV production (eq. \ref{eq:photon_production}), and the fact that all SFR variations are encapsulated in $\delta$ (eq. \ref{eq:eq:definition_variation}), a standard deviation in the SFR of $\sigma_\delta\simeq 0.24$ would induce an analogue s.d. in the UV magnitude of $\sigma_{\rm UV}\simeq 0.61$.
Such a variation is too small to explain the over-abundance of luminous galaxies seen by JWST at $z \simgt 10$. As shown by \citet{mason:2023} \citep[see also][]{shen:2023,munoz:2023}, a $\sigma_{\rm UV} \simeq 1.5$ ($\simeq 2$), i.e. $\simeq 3\times$ higher than what we find here, is required to reconcile models with data.
Alternatively, such requirement could be directly cast in terms of $\sigma_\delta$ and compared with the results in \citet{mirocha:2023}, which imply that $\pm 1$ dex scatter in the SFR is needed, i.e. $\simeq 4 \times$ the value found here. 

\section{Discussion}
Our result show that the over-abundance of super-early, luminous JWST galaxies cannot be easily explained by SFR stochasticity. Thus, it is necessary to explore different scenarios. 

Among the various possibilities, it has been suggested \citep{ferrara:2023} that radiation-driven outflows, which are expected to be very common among these luminous, compact objects \citep{fiore:2023,ziparo:2023}, effectively clear the dust making the galaxies more luminous. Unfortunately, at present, this effect is only coarsely modelled in \code{SERRA} and other similar simulations.

Alternatively, modifications to the $\Lambda$CDM cosmology have been also considered \citep{Boylan23, Gong23, Haslbauer22, parashari:2023}. Although interesting, a more thorough exploration of the implications of a different matter power spectrum on the properties and abundance of low mass galaxies ($M_\star\lsim 10^7\msun$), which show no tension with JWST data at the moment \citep[][]{mccaffrey:2023}, is necessary to draw firm conclusions. Particular care should be taken in solving the JWST over-abundance problem by modifying the $\Lambda$CDM power spectrum, since such a change can induce large tensions at lower $z$ \citep{gouttenoire:2023,sabti:2023}.

Finally, the feedback-free starburst scenario \citep[e.g.][]{dekel:2023} remains an -- albeit extreme -- intriguing alternative. However, the implications of such a scenario (globular cluster formation, merging of intermediate mass black holes, consequences for the reionization history) are yet to be explored.
%

\section{Summary}

We have analysed stochastic time variations $\delta(t)$ of the SFR in high-$z$ galaxies, by using the growth histories of 245 $z=7.7$ galaxies with stellar mass $5\times 10^{6}\lsim M_\star/\msun\lsim 5\times 10^{10}$ from the \code{SERRA} simulation suite \citep{pallottini:2022}.
After fitting the average star formation history, $\langle {\rm SFR} \rangle$, for each galaxy, the variation is quantified as $\delta(t) \equiv \log [\rm SFR/\langle {\rm SFR} \rangle]$. The main results are:  
\begin{itemize}
    \item[$\bullet$] The variation $\delta(t)$ is independent of $M_\star$, and is distributed as a zero-mean Gaussian with standard deviation $\sigma_\delta\simeq 0.24$.
    \item[$\bullet$] $\delta(t)$ is periodic on timescales that increase with $M_\star$: $t_\delta \sim (9, 50, 100)\, \rm Myr$ for $M_\star \sim  (0.1, 1, 5)\times 10^9\, \msun$, respectively. Such modulations for low, intermediate and high stellar mass are induced by (i) photoevaporation of molecular hydrogen, (ii) SN explosions and (iii) cosmic accretion. Feedback (either radiative and/or mechanical) regulation is important in the whole mass range;  cosmic accretion becomes the dominant variability source for $\simeq 5 \times 10^9\msun$ galaxies.
    \item[$\bullet$] SFR variations induce analogue $\rm UV$ magnitude variations with standard deviation $\sigma_{\rm UV}\simeq 0.61$. Such amplitude falls short by $\simeq 3\times$ \citep[][]{shen:2023} or $\simeq 4\times$ \citep{mirocha:2023} to explain the over-abundance of luminous, $z\gsim 10$ galaxies seen by JWST. Such over-abundance is instead more readily explained by models in which radiation-driven outflows efficiently clear the dust from these super-early systems (e.g. \citealt{ferrara:2023}).
\end{itemize}

\begin{acknowledgements}
AF acknowledges support from the ERC Advanced Grant INTERSTELLAR H2020/740120 (PI: Ferrara).
Any dissemination of results must indicate that it reflects only the author's view and that the Commission is not responsible for any use that may be made of the information it contains.
Partial support (AF) from the Carl Friedrich von Siemens-Forschungspreis der Alexander von Humboldt-Stiftung Research Award is kindly acknowledged.
We acknowledge the CINECA award under the ISCRA initiative, for the availability of high performance computing resources and support from the Class B project SERRA HP10BPUZ8F (PI: Pallottini).
We gratefully acknowledge computational resources of the Center for High Performance Computing (CHPC) at SNS.
AP would like to thank R. Poggiani for a discussion on the Lomb-Scargle periodogram of a few years ago.
We acknowledge usage of the Python programming language \citep{python2,python3}, Astropy \citep{astropy}, Cython \citep{cython}, Matplotlib \citep{matplotlib}, Numba \citep{numba}, NumPy \citep{numpy}, \code{pynbody} \citep{pynbody}, and SciPy \citep{scipy}.
\end{acknowledgements}

\bibliographystyle{stile/aa_url}
\bibliography{master,codes}

\end{document}